\documentclass[a4paper, 12pt]{article}
\pdfoutput=1
\usepackage{cite}
\usepackage{epsfig,amssymb,bbm,euscript,xspace,xcolor,setspace,amsmath,mathtools,empheq,amsthm,graphicx,mathrsfs,float}
\usepackage{hyperref}
\usepackage{bbding}
\usepackage{xcolor}
\hypersetup{
    colorlinks=true,
    linkcolor=black,
    citecolor=blue
    } 
\usepackage{caption,subcaption} 
\usepackage[belowskip=-20pt,aboveskip=0pt]{caption}
\usepackage[skip=10pt]{caption} % example skip set to 2pt
\usepackage{comment}
\usepackage{wrapfig}
\usepackage[numbers,sort&compress]{natbib}
\usepackage{cite}
\newcommand\blfootnote[1]{%
  \begingroup
  \renewcommand\thefootnote{}\footnote{#1}%
  \addtocounter{footnote}{-1}%
  \endgroup
}

\makeatother

\begin{document}
\newcommand{\jmh}[1]{ \textcolor{blue}{#1}}
\newcommand{\jmhadd}[1]{ \textcolor{red}{[#1]}}
\newcommand{\pr}[1]{ \textcolor{red}{#1}}
\hspace{370pt}{MPP-2024-122}

\thispagestyle{empty}

\baselineskip 10pt
\vspace*{1.0ex}
\begin{center}
{\Large \bf Positivity properties of scattering amplitudes} 
\end{center}

%\vskip .25cm
%\medskip

\vspace*{2.0ex}

\baselineskip=18pt

\centerline{\large \rm  Johannes Henn$^{s}$\blfootnote{$^{s}$henn@mpp.mpg.de}, Prashanth Raman$^{t}$ \blfootnote{$^{t}$praman@mpp.mpg.de}}
\vspace*{4.0ex}

\centerline{\it  Max-Planck-Institut f\"{u}r Physik, %Werner-Heisenberg-Institut, 
Boltzmannstr. 8,
%}
%\centerline{\it~ 
85748 Garching, Germany.}

\vspace*{0.1ex}
\centerline{\it ~} 
%\vspace*{5.0ex}
%\centerline{\bf Abstract} \bigskip
We investigate positivity properties in quantum field theory (QFT).
{We provide evidence, and in some case proofs, that many building blocks of scattering amplitudes, and in some cases the full amplitudes,}
satisfy an infinite number of positivity conditions: the functions, as well as all their signed derivatives, are non-negative in a specified kinematic region. Such functions are known as completely monotonic (CM) in the mathematics literature. A powerful way to certify complete monotonicity is via integral representations. We thus show that it applies to {planar and} non-planar {Feynman} integrals possessing a Euclidean region,
{as well as to certain Euler integrals relevant to cosmological correlators and stringy integrals.}
{This implies in particular that many basic building blocks appearing in perturbation theory, such as master integrals, can be chosen to be completely monotone}.
We also discuss two pathways for showing complete monotonicity for full amplitudes. One is related to properties of the analytic S-matrix. The other one is a close connection between the CM property and Positive Geometry.
Motivated by this, 
we investigate positivity properties in planar maximally supersymmetric Yang-Mills theory.
We present evidence, based on known analytic multi-loop results, that the CM property extends to several physical quantities in this theory.
This includes the (suitably normalized) finite remainder function of the six-particle maximally-helicity-violating (MHV) amplitude, four-point scattering amplitudes on the Coulomb branch, four-point correlation functions, as well as the angle-dependent cusp anomalous dimension.
Our findings are however not limited to supersymmetric theories.
It is shown that the CM property holds for the QCD and QED cusp anomalous dimensions, to three and four loops, respectively.
We comment on open questions, and on possible numerical applications of complete monotonicity.
\vfill \eject
\baselineskip=18pt

\tableofcontents
\onehalfspacing

\section{Introduction}

It is well known that certain observables in quantum field theory (QFT) enjoy positivity properties that can be derived from first principles, such as unitarity and analyticity/locality.
Examples are bounds on effective field theory coefficients \cite{Arkani-Hamed:2020blm,Bellazzini:2020cot}, or the conformal bootstrap \cite{Simmons-Duffin:2016gjk}.
In the Positive Geometry program, 
%initiated in refs. \cite{Hodges:2009hk, Arkani-Hamed:2013jha}, see \cite{Herrmann_2022} for a review,
reviewed in \cite{Herrmann_2022}, positivity is the input, and the output, is an alternative definition of scattering amplitudes.

{First hints for positivity properties of scattering amplitudes were found in the beautiful paper \cite{Hodges:2009hk}.
In this paper, tree-level next-to-maximally-helicity-violating (NMHV) Yang-Mills 
were expressed as the volumes of certain polytopes in momentum twistor space.
Subsequently, this observation was extended to one-loop MHV integrands \cite{Arkani-Hamed:2010wgm}.}

Motivated by this discovery, the authors of \cite{Arkani-Hamed:2013jha} introduced the Amplituhedron, which describes both tree-level amplitudes and planar loop-level integrands in sYM theory have an interpretation as the (conjecturally, unique) canonical form of a certain geometry associated to kinematic space (and in general, to certain auxiliary variables). This advance provides a definition of those objects that is completely different from usual field theory approaches. Indeed, field theory properties such as unitarity, locality, or the structure of infrared divergences are shown to emerge from this definition.

%The Positive Geometry program aims at a novel, geometric reformulation of quantum field theory. In this new framework, a given theory has a putative positive geometry associated with it, and scattering amplitudes are to be thought of as differential forms rather than as functions. Unitarity and locality emerge from the geometry rather than being physical inputs. The canonical differential form associated with this geometry corresponds to the scattering amplitude (or, at loop level, to their integrand).

%This framework has been established for all loop amplitudes in planar $\mathcal{N}=4$ SYM theory and $Tr(\phi^3)$ theory where the corresponding positive geometries are the amplituhedron and surfacehedron respectively.
%For scattering amplitudes, the starting point is a geometric object defined by the kinematic data of the scattered particles. 

In cases where the geometry is a polytope, it is understood that the canonical form, i.e. the integrand of the scattering amplitude, has an interpretation as the volume form of the dual polytope. 
An important open question is whether beyond these simple cases, the existence of a dual Amplituhedron and an associated volume interpretation for amplitudes is conjectural. If true, this could fit naturally into the AdS/CFT picture for scattering amplitudes, which itself has a volume interpretation at strong coupling \cite{Alday:2007hr}.

It is an interesting direction to explore whether integrated objects can also have nice positivity properties. 
This is a non-trivial expectation which would require that Minkowski integration contour preserves positivity and would be indicative of some intrinsic fundamental mechanism. However, for planar sYM evidence for positivity of a suitably defined infrared-finite part of six-particle amplitudes, when evaluating the external kinematics inside the tree Amplituhedron, was found in reds. \cite{Arkani-Hamed:2014dca, Dixon:2016apl}. 
Furthermore, in \cite{Dixon:2016apl} the authors provide evidence for not just positivity but also monotonicity (negativity of the first derivative) in a radial direction. In \cite{ Chicherin:2022zxo,Chicherin:2024hes} it was shown that the integrated two-loop result for the null pentagonal Wilson loop with a Lagrangian insertion (which is closely related to scattering amplitudes) also exhibits positivity properties inside the Amplituhedron. 

Given these remarkable observations,
while the exact mechanism of how integrated objects inherit the positivity of integrands is not fully understood, one can ask: are there other non-trivial consequences of the underlying geometric picture, apart from positivity of the result?
The following simple argument, inspired by \cite{Arkani-Hamed:2014dca}, provides hints as to why one might anticipate further properties.
Consider an amplitude that is given by the volume ${\rm Vol}(A^{*})$, where $A$ is the associated geometry, and $A^{*}$ denotes its dual. A simple fact about duality is that if $ A \subset B$ then $B^{*} \subset A^{*}$. So, if we increase the size of the geometric object, then the size of the dual decreases.
 This suggests that in addition to positivity, one may expect negativity of (suitably defined) derivatives.

In this work we go further and report on  an infinite number of positivity conditions involving scattering amplitudes and all their higher order derivatives. The relevant property is known as {\it complete monotonicity} in mathematics \cite{widder1941laplace}. Our main motivation for investigating these surprising properties is their close connection to positive geometries. As we will see, the characterization of this property is fundamentally connected to the dual volume interpretation. 
This will be totally manifest in cases where the geometry is a polytope. One can hope that these insights will teach us something new about the dual volume picture in cases beyond polytopes. 

As this is the first paper on this subject, we limit ourselves to introducing the notion of complete monotonicity in the context of scattering amplitudes, present heuristic evidence for its relevance to several quantities both in sYM and beyond, and outline directions that we find promising in view of eventually proving the CM property a priori.

As an added bonus we shall also argue why some objects closely related to positive geometry such as the cusp anamalous dimension, Euler-type integrals, and scalar Feynman intgegrals also share these properties in their corresponding Euclidean regions directly via some integral representations.

%is two fold. Firstly, we hope that this provides a useful first step towards investigating positivity of integrated objects. }  \pr{SAY MORE ABOUT WHY}   

{The outline of this paper is as follows.
We begin by reviewing the definition of completely monotonic functions in section \ref{sec:CMintro}, with examples relevant to perturbative quantum field theory. 
In section \ref{sec:abundance}, we show how complete monotonicity can be derived from suitable integral representations.
We do so for large classes of integrals that are building blocks of QFT computations, such as planar Feynman integrals. Moreover, we discuss how complete monotonicity can be derived from dispersive representations of amplitudes,
if the latter satisfy certain analyticity and unitarity conditions.
Next, we discuss how Choquet's theorem provides a possible explanation of the CM property in terms of Positive Geometry.
{Subsection \ref{subsec:discussion} provides a dsicussion of subtle features of complete monotonicity}.
We then turn, to observables related to positive geometry. We present evidence of complete monotonicity of six-particle sYM amplitudes, and, in section \ref{sec:moreevidence}, for several further quantities, such as the cusp anomalous dimension. We present our summary and discuss open questions in section \ref{sec:summary}.}

\section{Completely monotonic functions}
\label{sec:CMintro}

Let us briefly introduce the relevant mathematical concepts.
A completely monotonic function $f(x)$, defined in a region $R \subset  \mathbb{R}$, satisfies an infinite number of positivity conditions \cite{widder1941laplace},
\begin{align}\label{eq:def:monotonex}
(- \partial_x)^n f(x) \ge 0 \,,\quad {\rm for} \quad n\ge 0\,,~~~\forall x \in R  \,.
\end{align}
 In other words, the function and all of its signed derivatives are positive. In particular, this means that such functions are non-negative, monotonically decreasing, and convex. This leads to typical {\it{convex} shapes of the functions},  cf. Figs.~\ref{fig:Efunction}, \ref{fig:cusp} below for examples.
  The fact that such functions have `completely boring' plots may be a virtue when it comes to numerical predictions in situation where full analytic results are not available, or that are time-intensive to evaluate numerically.

Let us quickly review useful features of CM functions. They are closed under multiplication and under taking convex sums: given two CM functions $f_1, f_2$, $f_1 \times f_2$ and $c_1 f_1 + c_2 f_2$ for $c_1 \ge 0,c_2 \ge 0$, are again CM functions. Likewise, one may generate further CM functions by taking (signed) derivatives, or by integrating (with a suitable choice of boundary constant).

  Let us give simple examples of CM functions, for which the CM property can be verified by differentiation. $1/(x+\alpha)$ ~with~ $\alpha >0$, and $\beta^x$ with $0<\beta<1$, which are CM in  $x \in(0,\infty)$. Another example is $-\log{x}$, which is CM in  $x \in(0,1)$. 

The Bernstein-Hausdorff-Widder (BHW) theorem \cite{widder1941laplace} states that a function $f(x)$ is completely monotonic 
on $x \in (0,\infty)$ if and only if it is the Laplace transform of a non-negative function $\mu(t)$, i.e.,
\begin{align}\label{Laplace-representation}
f(x) = \int_0^{\infty} e^{-t x} \mu(t) dt \,.
\end{align} 
An example is 
\begin{align} \label{logexample}
f(x)=\frac{\log x}{x-1}\,, \quad {\rm with} \quad \mu(t) = \int_0^{\infty} \frac{e^{-t y}}{y+1}dy\,,
\end{align}
which is manifestly non-negative. 

Note that the CM property depends on both the choice of variable and region. In general, composition of functions {\it does not} preserve it. However, if $f$ is completely monotonic and $g$ is absolutely monotonic (i.e., itself and all of its derivatives are non-negative), then $g(f)$ is CM. {In particular, if $f(x)$ is CM, then $e^{f(x)}$ is also CM. Another composition which preserves the property is the following: if $f$ is CM and $h$ is non-negative with a derivative that is CM, then $h(f)$ is CM. From this we can see that if $f(x),g(x)$ are CM, then so is $f\left(\int_0^x g(t) dt \right)$.}

Moreover, complete monotonicity is preserved under taking limits. {I.e., if we have a sequence of functions $\{f_n(x)|~ f_n ~{\rm is~ CM ~in~ R_n}\}$ for all $n \geq 0$ then the pointwise limit of the sequence if it exists is also CM. For instance, $f_n(x) = \left(1-\tfrac{x}{n}\right)^n$ is CM for $R_n= (0,n)$ and taking the limit we obtain $e^{-x}$, which is  indeed CM in $x \in (0,\infty)$.}

The multi-variable version of CM functions satisfies \cite{10.1007/BFb0099850}
\begin{align}\label{cmmulti}
(-\partial_{x_1})^{m_1} \ldots (-\partial_{x_n})^{m_n} f(x_1, \ldots , x_n) \ge 0 \,.
\end{align}
A first example is a linear function of $x_i$, with non-negative coefficients $c_i$, raised to a negative power $\alpha <0$,
\begin{align}\label{eq:linearnegativepower}
   f(x_1, \ldots , x_n) = {(c_1 x_1 + \ldots c_n x_n)^{\alpha}}\,.
\end{align}
The case of non-linear polynomials raised to negative powers is more complicated and involves hyperbolic polynomials \cite{Scott_2014,kozhasov2019positivity}.

As a two-variable example involving polylogarithms, the 
following function appears in a finite seven-point one-loop integral \cite{Arkani-Hamed:2010zjl},
\begin{align}
\begin{split}
\Psi^{(1)}(x_1,x_2) =& \, {\rm Li}_{2}(1-x_1) +{\rm Li}_{2}(1-x_2)   + \log x_1 \log x_2 - \pi^2/6 %\zeta_2 
\,.
\end{split}
\end{align}
To see that this is CM, it is useful to consider
\begin{align}\label{defPsi}
f(x_1,x_2) =& \frac{\Psi^{(1)}(x_1,x_2)}{1-x_1 - x_2}\,,
\end{align}
which has 
the dispersive integral representation,
\begin{align}
f(x_1,x_2) =  \int_0^{\infty}\int_0^{\infty}  \tfrac{dy_1 dy_2}{(x_1 + y_1) (x_2 + y_2) (1+y_1+y_2) } \, .
\end{align}
From this equation it is manifest that $f$ is CM on $x_1 >0, x_2 >0$.
Moreover, thanks to the product rule, and in view of eq. (\ref{defPsi}), we can deduce that $\Psi^{(1)}$ is CM, as long as $x_1+x_2 \le 1$.

Multi-variable CM functions are characterized by a generalized version of the BHW theorem  
due to
Choquet \cite{Choquet1954}. 
To state this theorem, we introduce the notion of CM functions on convex cones. 
 Let $V$ denote a finite-dimensional real vector space and $C$ be an open convex cone in $V$, i.e, a convex subset of $V$ such that if $x\in C$ then $\lambda ~ x \in C$ for all $\lambda>0$.  The closed dual cone $C^{*}$ is a subset of the dual vector space $V^{*}$ and defined as $C^{*}=\{y\in V^{*}| \langle y,x\rangle \geq 0 ~\forall~ x\in C  \}$, where $\langle y,x\rangle =\sum_{i} y_i x_i$ denotes the Euclidean inner product.
 
A function $f:C \rightarrow \mathbb{R}$ is completely monotone if it is $\mathbb{C}^{\infty}$- differentiable and satisfies 
\begin{equation} \label{cmcone}
(-1)^k D_{v_1}\cdots D_{v_k} f(x) \geq 0\,,\quad \forall x \in C
\end{equation}
where $D_v$ is the directional derivative along the vector $v$ with $k \in \mathbb{N}$, and for any set of vectors $v_1,\cdots,v_k \in C$. %

As an example, consider
$f(x,y) = \log y/x$ on the cone $C=\{(x,y) | 0<x<y \}$. 
We see that the cone $C$ is bounded by the vectors $v_1= (0,1)$ and $v_2=(1,1)$.
%corresponding to $x=0$ and $x=y$. 
Any vector in $C$ can be expressed as a positive combination of $v_1$ and $v_2$. We can verify the CM property of eq. (\ref{cmcone}) by taking directional derivatives along these rays.
%which result in either $\frac{(n-1)!}{x^n}$, $(n-1)!\left(\frac{1}{x^n}-\frac{1}{z^n}\right)$ or $\frac{(n-2)!}{x^{n-1}}\left(\frac{1}{x}-\frac{1}{z}\right)$ which are all manifestly positive inside $C$ thereby showing this function is CM.

A characterization of multi-variables CM functions is provided by the 
%following result.  
{\bf Bernstein-Hausdorff-Widder-Choquet} theorem: Let $f$ be a real-valued function on the open cone $C \subset \mathbb{R}^n$. Then $f$ is completely monotone if and only if it is the Laplace transform of positive measure $\mu$ supported on the dual cone $C^{*}$ 
\begin{equation} \label{choquet1}
    f(x)= \int_{C^{*}}  e^{-\langle y, x\rangle} \mu(y) dy \,.
\end{equation}
We note that when $V=\mathbb{R}^n$ and $C=\mathbb{R}^n_{+}$ then eq. \eqref{cmcone} reduces to eq. \eqref{cmmulti}. We also note that in this case the dual cone $C^{*}=\mathbb{R}_{+}^n$ and eq. \eqref{choquet1} becomes a generalization of HBW with the integral becoming the standard multidimensional Laplace transform.

As we have seen in this section CM implies positivity in the CM region. Note that only a small subset of positive functions are CM.  A trivial class of counter examples are polynomials with positive coefficients. A less trivial example is the function 
\begin{equation}
f(x,y)= \frac{ A \log^2{\tfrac{x}{y}}+ \pi^2}{x ~y} 
%= \int_{\mathbb{R}_{+}^2} dx dy e^{-x u -y v}  \left( A \log^2{\tfrac{u}{v}}+ \pi^2 \left(1-\tfrac{A}{3} \right)\right) 
\,,
\end{equation}
which for $x>0, y>0$ is positive for $ 0 \le A$,  but CM only for $0\leq A \leq 1/3$, as can be seen from computing its  inverse Laplace transform.

\section{Proof of complete monotonicity from integral representations}
\label{sec:abundance}

We show that CM functions show up in numerous QFT building blocks.
This is easily seen from suitable integral representations, as we discuss presently.

\subsection{Scalar Feynman integrals}
\label{sec:feynman}

An important case of CM functions is that of scalar {Feynman integrals}, subject to conditions that we explain presently. Consider the Feynman parametrization for a graph G with $L$ loops in $D=4-2 \epsilon$ dimensions (see e.g. \cite{Weinzierl:2022eaz}),
\begin{equation}\label{defFeymanIntegral}
    I(x)=
        \frac{\Gamma(\sum_{i}\nu_i - LD/2)}{\prod_{i}\Gamma(\nu_i)}\int_{\alpha_i \geq0} \frac{\prod_{i} d \alpha_i \alpha_i^{\nu_i-1}}{{\rm{GL}(1)}}  \frac{U(\alpha)^{\sum_i \nu_i-(L+1) D/2}}{F(\alpha,x)^{\sum_i \nu_i-L D/2}}\,,
\end{equation}
where $U,F$ are the following graph polynomials 
\begin{align}
    U(\alpha)&= \sum_{T\in T_1}\prod_{e_i\notin T} \alpha_i \,, \\
    F(\alpha,x)&= \sum_{(T,R)\in T_2}
    \left(\prod_{e_i \notin(T,R)} \alpha_i \right) (-s_{T,R})+U(\alpha) \sum_{i=1}^n \alpha_i m_i^2\,,
    \end{align}
where, $T_1,T_2$ are the spanning trees and two-forests of the graph G, respectively, {and $s_{T,R}= \left(\sum_{e_j \notin (T,R)} q_j \right)^2$. %denotes the square of the momenta through the deleted lines of $(T,R)$. 
We have collectively denoted the kinematic variables as $x= \{-s_{T,R},m_i^2\}.$   }

Let us now assume that the powers $\nu_i$ and the dimension $D$ are such that the integrations in eq. (\ref{defFeymanIntegral}) converge.
Notice that the kinematic variables $x$ enter eq. (\ref{defFeymanIntegral}) via $F$ only, which in turn is linear in them.
Moreover, their coefficients are non-negative, due to $\alpha_i \ge 0$.
This implies, as in eq. (\ref{eq:linearnegativepower}) that the integrand is a CM function of $x$ for $x > 0$.
The CM property is preserved by carrying out the integrations. From this it follows that $I(x)$ is a CM function.
Another way of seeing this is by differentiating under the integral sign.

As an example, the one-loop massive bubble integral in two dimensions has the following Feynman representation,
\begin{align}\label{eq:bubblemassive}
f(x_1,x_2) = \int_0^{\infty} \frac{d \alpha_1 d\alpha_2}{{\rm GL}(1)} \frac{1}{x_1 \alpha_1 \alpha_2 + x_2 (\alpha_1 + \alpha_2)^2}\,,
\end{align}
where $x_1 = -P^2\,, x_2 = m^2$. It is a CM function of $x_1,x_2$ for $x_1 , x_2 > 0$, as can be seen e.g. by differentiating under the integral in eq. (\ref{eq:bubblemassive}).

There is a subtlety in the previous argument, as it assumes that all $x$ variables are independent and a region $x>0$ exists. Due to kinematic constraints, such as momentum conservation, this may not always be the case.
A case in point is massless on-shell four-particle scattering, where $x=\{-s,-t,-u\}$, which satisfy $s+t+u=0$, which does not allow $x>0$. This is not the end of the story, however.
Firstly, for planar Feynman diagrams, the $u$-channel variables are absent from $F$, so that e.g. for on-shell planar four-particle kinematics, simply choosing $-s>0, -t>0$ satisfies the CM requirements, at any loop order. In general, the region where $F >0$ is called the Euclidean region.
Secondly, even in the non-planar case, there may exist a Euclidean region. In partcular, this is the case for sufficiently general kinematics.
For example, the four-point non-planar integrals considered in ref. \cite{Henn:2013nsa} have a Euclidean region if one external leg is off-shell, but not otherwise.
{A sufficient criterion for the existence of a Euclidean region was given in reference \cite{Mizera:2021ujs}}.
The general question of determining the Euclidean region is an interesting open question.

Let us comment on the requirement of convergence of the integral.  
Consider for example the scalar four-dimensional one-loop box integral in $D=4- 2 \epsilon$ dimensions. It can be written as 
(see e.g. \cite{Smirnov:2012gma})
\begin{align}\label{eq:boxdimreg}
\frac{\Gamma^2(-\epsilon) \Gamma(2+\epsilon)}{\Gamma(-2 \epsilon)} \int_0^1 \int_0^1 
[(-s) y_1 y_2 + (-t) (1-y_1)(1-y_2)]^{-2-\epsilon} dy_1 dy_2 \,.
\end{align}
This is indeed CM in $(-s),(-t)$ and for $-2<\epsilon<0$, 
where the integral is ultraviolet and infrared finite.
In practice, one is often interested in a Laurent expansion as $\epsilon \to 0$. In this case, the leading term is CM, but coefficients of individual powers are in general not, as they correspond to a subtraction of the leading term(s).
But one may derive useful CM constraints via integration-by-parts and/or dimension-shift relations \cite{Chetyrkin:1981qh,Tarasov:1996br}. For example, it is well-known that the finite part of the four-dimensional one-loop box integral can be viewed as a finite, six-dimensional box integral.
Indeed, evaluating eq. (\ref{eq:boxdimreg}) for $\epsilon=-1$ gives
\begin{align}\label{resultbox6d}
-\frac{1}{2} \frac{\log^2(t/s) + \pi^2 }{s+t} \,,
\end{align}
and one may check that this is CM.
This is related to the discussion in section \ref{sec:sYM}, where we discuss complete monotonicity of suitably defined finite parts of scattering amplitudes in maximally supersymmetric Yang-Mills theory.

\subsection{Euler-type integrals}
\label{sec:euler}

We can use integral representations to ascertain the CM property for further relevant classes of quantum field theory objects.
An example are the integral representations for {\it cosmological correlators} discussed in \cite{Arkani-Hamed:2023kig}. E.g.
\begin{align} \label{eq:FLRW}
\begin{split}
\Psi_{\rm FLRW} \propto & \int_0^\infty  
 \frac{dx_1 dx_2  (x_1 x_2)^{\epsilon}}{(X_1+X_2 +x_1 +x_2)
 (
X_1 + x_1 +Y)(X_2 + x_2 + Y)} \,,
\end{split}
\end{align}
represents the contribution of a scalar tree-level diagram to the four-point wavefunction coefficient in a general Friedmann–Lemaître-Robertson–Walker (FLRW) spacetime.
We see that $\Psi_{\rm FLRW}$ in eq. (\ref{eq:FLRW}) is CM for $X_1, X_2, Y >0$.

Another example is the famous Veneziano formula,
\begin{align}
\frac{\Gamma(-\alpha' s) \Gamma(-\alpha' t)}{\Gamma(-\alpha' s- \alpha't)} = \int_0^1 \frac{dy}{y (1-y)} y^{-\alpha' s} (1-y)^{-\alpha' t} \,.
\end{align}
One can see that this is CM in $s,t$ for $s,t<0$ by noticing that a derivative in $-s$ (or $-t$)  inserts an extra factor of $\log y$ (or $\log(1-y)$) into the integrand, which is uniformly negative in the integration domain. 
This analysis can be extended to more general {\it stringy canonical forms} by virtue of the $u$-representation, cf. section 9.4 of  
ref. \cite{Arkani-Hamed:2019mrd},
 \begin{align}
\mathcal{I}(S)=(\alpha^{\prime})^d\int_{\mathbb{R}_{+}^d} \prod_{i=1}^d \frac{d x_i}{x_i} \prod_A u_A^{\alpha' F_A}\,.
\end{align}
Importantly, in this formula, $0<u_A<1$,  which makes it manifest that these integrals are CM in the interior of the Newton polytope defined by $F_A>0 $, which is where the integrals converge \cite{2011arXiv1103.6273B}. 
This is closely related to positivity certificates of Euler-type integrals discussed in reference \cite{kozhasov2019positivity}.

\subsection{Dispersive integrals}
\label{sec:subsectionanalytcSmatrix}

There is a close connection between complete monotonicity and {\it dispersion relations}. 

For example, consider an unsubtracted dispersion relation,
\begin{align}\label{dispersive1}
A(s)=& \int_{4 m^2}^{\infty}ds'\frac{1}{(s'-s)} F(s') \,,
\end{align}
where $F(s)$ is the discontinuity of $A(s)$ for $s>4m^2$. 
Using the Schwinger trick $1/(s'-s) = \int_0^\infty e^{-t (s'-s)} dt$, this can be rewritten as eq. (\ref{Laplace-representation}), with $x=-s$ and 
\begin{align}\label{mudispersive2}
\mu(t) = \int_{4m^2}^\infty e^{-t s'} F(s') ds' \,.
\end{align}
From eq. (\ref{dispersive1}) we see that $A$ is a CM function of $x$, provided that $F(s')$ is non-negative. However, a weaker necessary condition is that $\mu(t)$ in eq. (\ref{mudispersive2}) is non-negative. 

This also turns out to be a useful trick to prove the CM property in certain cases. A case in point is the $ {\log{x}}/{(x-1)}$ example of eq. \eqref{logexample}, which has the dispersive representation (\ref{dispersive1}), with $m=0$ and $F(s')={1}/({1+s'})$. This readily gives the non-negative expression for $\mu(t)$ quoted in eq. \eqref{logexample}.

The above analysis can be generalized to Mandelstam representations \cite{Mandelstam:1959bc,Eden:1966dnq}. Very interestingly, for a two-to-two scattering amplitude $M(s,t;m^2)$ of identical particles of mass $m$, the authors of references \cite{Jin:1964zz,Martin:1965jj,Martin:1970jsp} were able to derive uniform sign properties for derivatives in $s$ or $t$ in a certain kinematic region.
More specifically, the assumptions made (see also \cite{Correia:2020xtr}) on the amplitude are: (1) linear unitarity, meaning that the imaginary part of the amplitude has positive partial wave coefficients; (2) Froissart-Martin bound on the Regge limit; (3) Crossing symmetry;
(4) Maximal analyticity.
Based on these assumptions, the authors of \cite{Jin:1964zz,Martin:1965jj,Martin:1970jsp} derived a twice-subtracted dispersion relation for the amplitude, whose kernel is positive in a certain kinematic region. 
It follows that
\begin{align} \label{pos1}
\frac{\partial^n}{\partial s^n} M(s,t; m^2) > &\, 0 \,,~~~{\rm for~}~~2m^2-t/2 \le s\le 4 m^2\,.
\end{align}
This shows that $M(s,t;m^2)$ is CM in $x=-s$ for fixed $t,m^2$, in the region indicated in eq. (\ref{pos1}), and similarly for $s \leftrightarrow t$.
It is an interesting open question to study the sign properties of the amplitude under mixed derivatives.

We note that there are interesting related findings of notions of positivity in the context of renormalization group flow \cite{Hartman:2023ccw}, and for forward amplitudes \cite{Hui:2023pxc}.
In the latter context, one may think of eq. (\ref{dispersive1}) as a representation of a forward two-to-two scattering amplitude, i.e. at $t=0$. In this case $F(s')$ is positive due to the optical theorem.

\subsection{Complete monotonicity and Positive Geometry}
\label{sec:choquet}

In this subsection we argue that CM properties arise naturally in the context of Positive Geometry. Let us being with {Choquet's theorem, cf. eq. (\ref{choquet1}). This provides useful guidance as to which variables, and in which region to expect complete monotonicity.
To illustrate this, consider the case of convex polytopes $A \in \mathcal{P}^m(\mathbb{R})$ in projective space, which in simple cases describe scattering amplitudes \cite{Hodges:2009hk}.
The canonical function (i.e. its canonical form, with the standard measure stripped off) of such a polytope is the volume of the dual polytope 
$A^{*}_Y$. 
The latter is defined by the facet inequalities $W\cdot Y>0$, for $Y \in A$ and for any $W \in A^{*}_Y$. The canonical function admits the following Laplace representation \cite{Arkani-Hamed:2017tmz},
\begin{equation}
{\Omega}(Y)=\frac{1}{m!} 
%\left( 
\int_{W \in A^{*}_Y}  e^{-W \cdot Y} d^{m+1}W
%\right)
\,.
\end{equation}
This proves that ${\Omega}(Y)$ is CM for any $Y \in A$ by Choquet's theorem, with measure $\mu=1$.
According to this theorem, the directional derivatives are to be taken along the extremal rays of the dual polytope $A^{*}_Y$.
This fact suggests to us a close connection between the CM property and dual geometries. 
This could help when looking for a dual geometry in cases beyond polytopes, such as the conjectured dual Amplituhedron \cite{Arkani-Hamed:2010wgm,Ferro:2015grk,Herrmann:2020qlt}.}
We leave these interesting questions to future work.

\subsection{Discussion} 
\label{subsec:discussion}

We have seen several examples of objects in QFT that are CM in this section. It is important to emphasize that this is a special property and we do not expect it hold in general. To make this point clear we would now like to provide a few non-examples.

We would like to stress that the CM property, if it holds, depends on the normalization. In other words, changing the latter may result in the property no longer being true.
As an example, consider the scalar six-dimensional box integral of eq. (\ref{resultbox6d}). 
If one normalizes this integral by its leading singularitiy, $-(s+t)$, one obtains
\begin{align}
\frac{1}{2}  \log^2 ({t}/{s}) + \frac{1}{2}  \pi^2 \,.
\end{align}
While this function is positive on $(-s,-t) \in \mathbb{R}^2$, it is not CM, which can be seen from the fact that is has a minimum at $t=s$.
Of course, many other explicit counter examples also exist, such as a generic Feynman integrals with non-trivial numerator, and the non-planar contribution to the correlation functions of Wilson loop with a Lagrangian insertion at three loops \cite{henn2019fourloopcuspanomalousdimension}, just to mention a few.

As a further comment, interesting QFT observables may contain Feynman integrals with loop-dependent numerator factors, which may spoil the CM property.
However, one can always choose a basis of scalar master integrals, thanks to integration-by-parts relations.   
Although the coefficients of such master integrals would in general not be sign-definite, having master integrals with nicer properties could be advantageous, for example for numerical evaluation.

In some cases however, the CM property even holds for the full amplitude, and not just for its building blocks. Indeed we just discussed possible explanations for this in subsections \ref{sec:subsectionanalytcSmatrix} and \ref{sec:choquet} (see also section \ref{sec:sYM}).

%Another subtle point is the choice of variables. 
%Complete monotonicity depends on the choice of variables. In other words, in general it is necessary to find the right variables in order to discover it. 
%\jmh{See section \ref{sec:moreevidence} for an example, where we report on our investigation of CM properties for two choices of variables, in the case of the cusp anomalous dimension. We found that while the CM property held in either variable up to two loops; starting from three loops, they failed to hold for one of the choices.}

Let us emphasize that, as mentioned below eq. (\ref{logexample}), complete monotonicity depends on the choice of variables and region.
This makes it rather non-trivial to search for CM properties, even if a rich set of perturbative data is available.
However, one may observe that in most discussions of scattering amplitudes, a small number of preferred variable choices show up, such as e.g. the masses and momenta of the particles, (ratios of) Mandelstam variables, or (momentum) twistor variables.
This suggests to us that a heuristic approach may be fruitful, where one checks for CM properties in some of these variables.
Moreover, if one assumes that the choice of variables should be independent of the loop order, then this becomes a rather non-trivial prediction that can be tested based on the rich perturbative data that is available for certain observables. We report on findings along these lines in sections \ref{sec:sYM} and \ref{sec:moreevidence}.

%Related to this we would also like to mention that 
As a final commment, in cases with a clear connection to Positive Geometry it may be possible to overcome these heuristic considerations, as Choquet's theorem provides both a natural set of variables and region for the quantity considered.

%Let us now imagine a quantity for which one suspects the existence of such a theorem, and hence CM properties, but where this has not been established. 

%In practice, we often have access to a rich set of perturbative results, which we would like to check for CM properties.
%But which variables should one use?

\section{Positivity of six-particle sYM amplitudes}
\label{sec:sYM}

As alluded to in the introduction and in subsection \ref{sec:choquet} above, an important motivation for studying integrated quantities from the perspective of complete monotonicity comes from Positive Geometry. 
The prime example of a positive geometry is the Amplituhedron, which determines the loop integrands in planar ${\cal N}=4$ super-Yang-Mills (sYM), which are rational
functions. These integrands have a volume interpretation, and they are positive within the Amplituhedron region \cite{Hodges:2009hk,Arkani-Hamed:2010wgm,Arkani-Hamed:2013jha}.
What happens when one integrates the integrand over Minkowski space?
The authors of \cite{Arkani-Hamed:2014dca} found evidence that the finite part of integrated amplitudes is also positive, when evaluating the external kinematics within the tree
Amplituhedron region.

Loop amplitudes have infrared divergences, and one may wonder whether putative positivity properties survive infrared subtraction. See reference \cite{Arkani-Hamed:2014dca} 
% pages 23-25 
for a discussion of this point.  
The choice of infrared subtraction scheme presents a subtlety.
{While a minimal subtraction of infrared divergences is possible, removing the (non-minimal) so-called Bern-Dixon-Smirnov (BDS) ansatz \cite{Bern:2005iz} has the advantage of leading to a dual-conformally-invariant answer \cite{Drummond:2007au}, which depends on three cross-ratios $u,v,w$ only. Furthermore, it was found in \cite{Caron-Huot:2016owq} that a further finite subtraction, called BDS-like, leads to a function that satisfies the Steinmann relations. In reference \cite{Dixon:2016apl} the same function, denoted by ${\cal{E}}(u,v,w)$ there, was found to have positivity properties up to four loops. It is therefore a natural object for us to study.}
Let us discuss this case in detail.
The six-particle tree MHV Amplituhedron region is given by 
\begin{align}
P_{\rm MHV}: {\Bigg\{ } \begin{split}
 & u>0, v>0, w>0, u+v+w<1, \\
 & (u+v+w-1)^2<4 u v w 
\end{split} \Bigg\} \,.
\end{align}
We expand quantities in planar sYM perturbatively in the Yang-Mills coupling $g_{\rm YM}$ using the following notation (and likewise for other quantities considered below),
\begin{align}
{\cal E}(u, v, w) = \sum_{L \ge 1} g^{2 L} {\cal E}^{(L)}(u, v, w)\,,
\end{align}
where $g^2 = g_{\rm YM}^2 N_{c}/(16 \pi^2)$.
The authors of reference \cite{Dixon:2016apl}  found that for $L \le 4$, $(-1)^L {\cal E}^{(L)}(u, v, w) \ge 0$, for kinematics in $P_{\rm MHV}$.
They showed this analytically in certain limits and on kinematic slices, and by
numerical evaluation for randomly chosen kinematic points in $P_{\rm MHV}$. They also found evidence of monotonicity in a double scaling limit. 

In this paper, we provide evidence that $(-1)^L {\cal E}^{(L)}(u, v, w)$ is CM in $P_{\rm MHV}$. We prove this analytically for $L=1,2$, and provide numerical evidence at $L=3,4$. 
%\pr{Say about choice of variables u,v,w and uniquenqness, naturalness etc } % JMH: Later.

We begin by proving complete monotonicity at one loop. We have \cite{Dixon:2016apl}
\begin{align}
-{\cal E}^{(1)}(u,v,w) =& f(u) + f(v) + f(w) \,,
\end{align}
with
\begin{align}
f(x) =& -{\rm Li}_{2}(1-1/x)\,.
\end{align}
Note that $P_{\rm MHV}$ implies $0<u,v,w<1$.
Therefore it is sufficient to prove that $f(x)$ is CM for $x \in (0,1)$.
We first note that $f(1)=0$. Therefore, if we can prove that $-\partial_x f(x)$ is CM, then the same property for $f(x)$ follows via integration.
To this end, we compute
\begin{align}
-\partial_x f(x) = \frac{1}{x} \times \frac{\log x}{x-1}\,.
\end{align}
The RHS of this equation is a product of CM functions, and is hence CM itself, which completes the proof.

At two loops, ${\cal E}$ is given by a weight four function \cite{Goncharov:2010jf,Dixon:2016apl}. 
We outline the proof that this function is CM below.
We employ a representation derived in \cite{Dixon:2011nj}, namely
\begin{align}
\begin{split}
    {\cal E}^{(2)}(u,v,w) =&  
    \tilde{r}(u)  + \tilde{r}(v) + \tilde{r}(w) + \Omega^{(2)}(u,v,w) 
   % \\ & 
  + \Omega^{(2)}(v,w,u) +\Omega^{(2)}(w,u,v)       
    \,.
\end{split}
\end{align}
Here $\Omega^{(2)}$ is a finite double pentagon integral \cite{Arkani-Hamed:2010pyv,Dixon:2011nj}, and $ \tilde{r}(u)$ is a weight-four function, 
expressed in terms of harmonic polylogarithms \cite{Gehrmann:2001pz}.
We prove that $\tilde{r}(u)$ is CM for $0<u<1$ by differentiation, and employing a dispersive representation. Next, we use 
\cite{Dixon:2011nj},
\begin{align}
{\Omega}^{(2)}(u,v,w) = \int_0^{w}H_{6}(u,v,t) dt  + \Psi^{(2)}(u,v)\,,
\end{align}
where $\Psi^{(2)}$ is a finite penta-box integral \cite{Drummond:2010cz}. The latter is seen to be CM for $u+v<1$ by considering its Feynman parametrization, in agreement with the CM property discussed above for its one-loop version $\Psi^{(1)}$.
Finally, $H_{6}$ is a scalar hexagon integral, which is CM, and hence so is ${\Omega}^{(2)}$. This completes the proof.

\begin{figure}[t]
   % \centering
   \begin{center}
    \includegraphics[width=80mm]{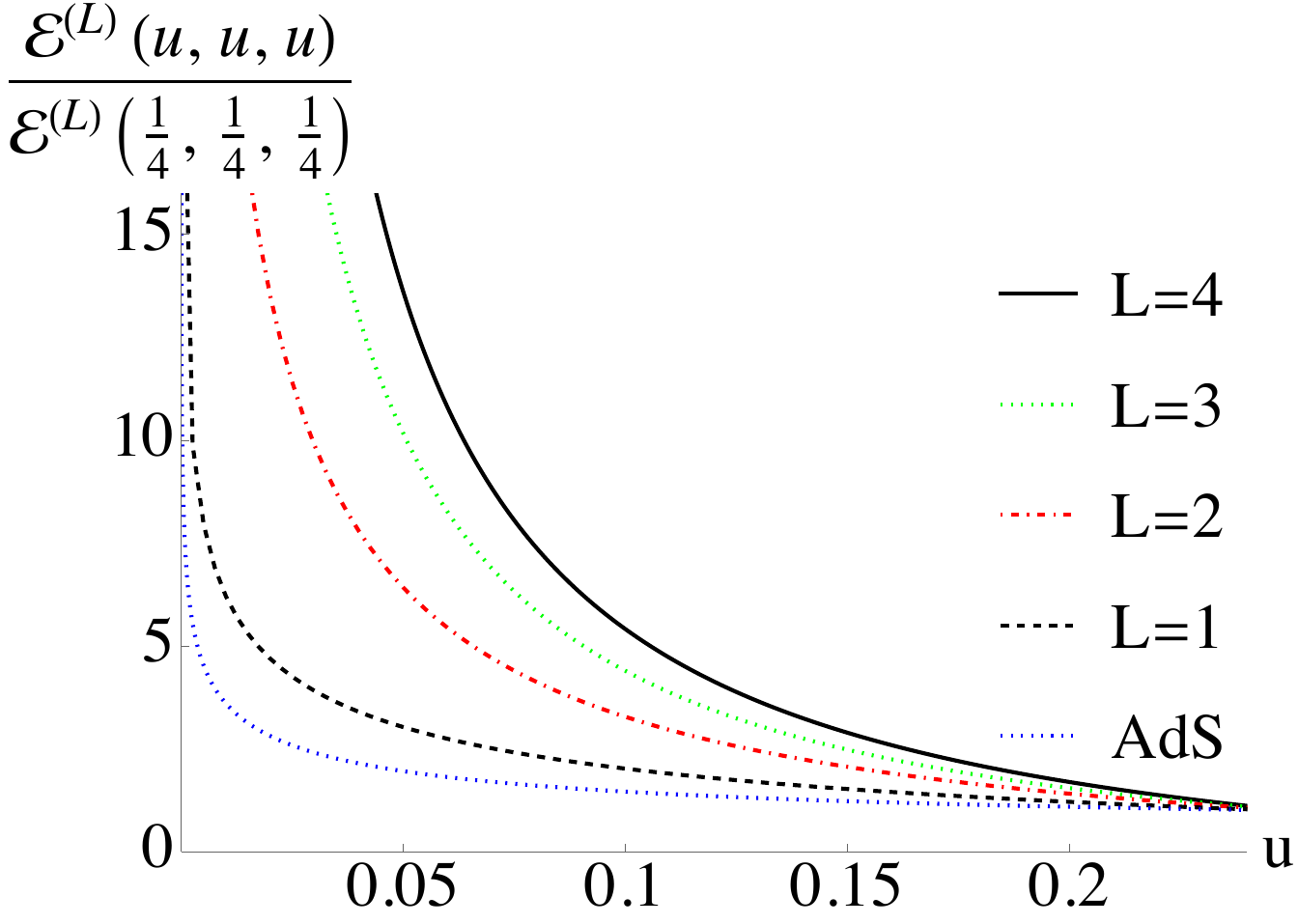}%
    \caption{Numerical evaluation of ${\cal E}^{(L)}(u,u,u)$ within the tree Amplituhedron region $0<u<1/4$. (A similar plot was shown, for larger values of $u$, in reference \cite{Caron-Huot:2016owq}.)
    From bottom to top: AdS (strong coupling) and $L=1,2,3,4$ loop results.
    The graphs are positive, monotonically decreasing, as well as convex, in agreement with complete monotonicity. 
    }
    \label{fig:Efunction}
    \end{center}
\end{figure}

At three and four loops, we performed the following numerical checks of complete monotonicity. Employing the analytic formulas computed in refs. \cite{Dixon:2013eka,Dixon:2014voa,Dixon6ptRepository4loop}, 
we evaluated ${\cal E}^{(L)}(u,v,w)$, as well as {all of its first and second derivatives, numerically for $10^4$ rational phase-space points within $P_{\rm MHV}$ that were generated using the {\tt{FindInstance}} command in {\tt{Mathematica 13.3}}. We evaluated the relevant Goncharov polylogarithms numerically at these points using the {\tt{HandyG}}  \cite{Naterop_2020} and {\tt{NumPolyLog}} \cite{ZhenjieLi} packages, with precision of 16 digits and accuracy of 10 digits, and similarly for the pre-factors.
At this stage we noticed that 
 at six and seven out of the $10^4$ phase-space points, for three and four loops respectively,
 had a signs that disagreed with our hypothesis. The phase-space points in question are close to the boundary of $P_{\rm MHV}$.
We suspect that this led to a loss of precision in the results. Indeed, when re-evaluating these points with increased precision of 50 digits, this issue disappeared. In the future, it may be interesting to develop a faster routine for such evaluations, also in view of higher-loop results.
  %We can provide the numerical data for the $10^4$ phase-space points %along with the list of these points where there was a loss of precision  
  %as ancillary text files.}
%Due to the complexity of the functions involved, 
%the numerical computation terminated within one week only for a subset of ${\cal{O}}(10^4)$ points. We used those points for the numerical check. 
%\jmh{We provide the points, as well as the numerical values obtained, in an ancillary file.}

Fig.~\ref{fig:Efunction} shows an interpolated plot for $u=v=w$, for which $P_{\rm MHV}$ becomes  $0<u<1/4$. The AdS curve corresponds to the strong coupling result \cite{Alday:2009dv,Basso:2014jfa,Basso:2020xts}, 
\begin{align}\log {\cal E}(u,u,u) =: g \, {\cal E}_{AdS} + {\cal O}(g^0)\,,
\end{align}
with \footnote{We thank Lance Dixon for insightful correspondence regarding the constants in eq. (\ref{EStrong}).} 
\begin{align}\label{EStrong}
{\cal E}_{AdS} = -\frac{3}{2 \pi} \log^2 \frac{1- \sqrt{1-4 u}}{1+ \sqrt{1-4 u}}  -\frac{11 \pi}{36} -\frac{\pi^2}{6}   
\,.    \end{align}
Note that $-{\cal E}_{AdS}$ is CM for $0<u<1/4$.

The functions shown Fig.~\ref{fig:Efunction} have shapes typical for CM functions, being in particular positive, decreasing, and convex. If complete monotonicity holds, plots in different kinematic slides within the Amplituhedron region would therefore look similar. This is rather remarkable if one recalls that the underlying analytic results for the six-particle amplitudes have a rich analytic structure, being expressed as iterated integrals of transcendental weight $2L$ at $L$ loops of integration kernels correspondong to an $A_{3}$ cluster algebra \cite{Goncharov:2010jf,Golden:2013xva}.

\section{Evidence of complete monotonicity of further physical quantities}
\label{sec:moreevidence}

We present evidence of the CM property for several further quantities in planar sYM.
In the first three cases, this evidence is based on numerical evaluation of the functions and their first two derivatives, as well as on numerical evaluation (whenever feasible) of the inverse Laplace transform. In the fourth case, the CM property can be proven.

{\it 1. Four-point Coulomb branch amplitudes} \cite{Alday:2009zm}, which depend on the kinematic variables $u=4 m^2/(-s), v=4 m^2/(-t)$.
We find numerical evidence that $(-1)^L {\cal M}^{(L)}(u,v)$ is a CM function of $u,v$, for $u >0,v >0$ using the available results \cite{Caron-Huot:2014lda} at $L=1,2,3$. 
Let us recall the small mass limit \cite{Alday:2009zm,Bruser:2018jnc}, in which
\begin{align}\label{Msmallmass}
{\cal M}(u,v;g)
\stackrel{m^2 \to 0}{\sim}e^{-  \frac{1}{2} \Gamma^{\infty}_{\rm cusp}(g) \log u \log v 
} \,,
\end{align}
where $\Gamma^{\infty}_{\rm cusp}(g) = 4 g^2 - 8 \zeta_2 g^4 \mp \ldots$ is the light-like cusp anomalous dimension \cite{Beisert:2006ez}.
We see that eq. (\ref{Msmallmass}) is consistent with the CM property as follows: First, the argument of the exponential is CM 
because $-\log u$ and $-\log v$ are CM functions for argument smaller than one, and because $(-1)^{L+1} \Gamma^{\infty\, (L)}_{\rm cusp}>0$. Second, the exponential of a CM function is also a CM function.

{\it 2. Four-point deformed Amplituhedron amplitudes}, computed to two loops in \cite{Arkani-Hamed:2023epq}.
We find that $(-1)^L {\cal M}^{(L)}(x,y)$ are CM functions of $x,y$, for $0<x<1, 0<y<1$, and $L=1,2$.

{\it 3. Four-point correlation functions of stress tensor multiplets.} We find numerical evidence for $x^2_{13} x_{24}^2 (-1)^L  F^{(L)}$ given in eq. (1.1) of \cite{Drummond:2013nda} being CM as functions of the cross-ratios $u,v$ given in eq. (1.11) of that paper, for $L=1,2$ (initially computed in \cite{Eden:2000mv,Bianchi:2000hn}). We leave the $L=3$ case for future work.

\begin{figure}[t]
%\centering
 \begin{center}
    \includegraphics
    [width=85mm]
    %[scale=0.3]
    {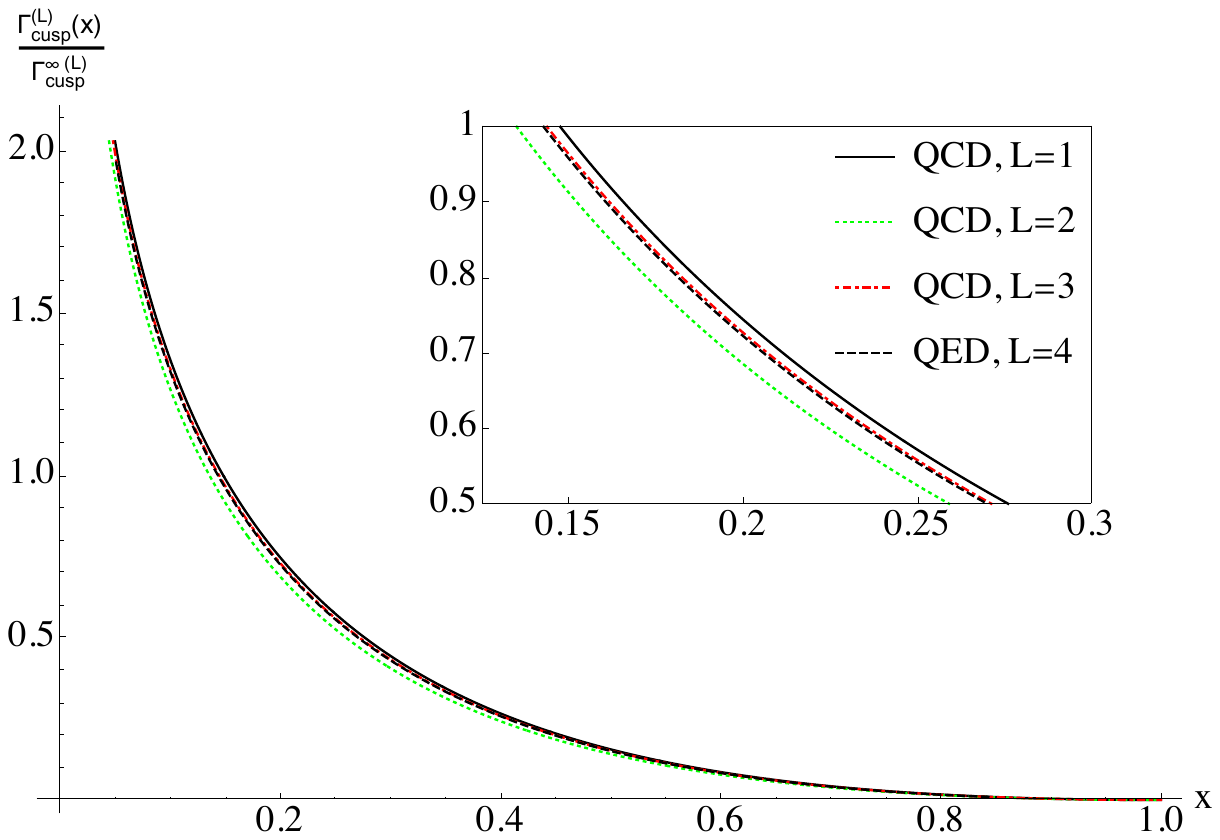}
       \caption{$L$-loop corrections to the angle-dependent cusp anomalous dimension in the Euclidean region $x \in (0,1)$, for the known QCD and QED cases.
       The inset shows a part of the plot in magnification. 
       }
    \label{fig:cusp}
    \end{center}
\end{figure}
%\vspace*{0.2 cm}

{\it 4. The angle-dependent cusp anomalous dimension}, up to four loops, cf. eq. (5.3) of \cite{Henn:2013wfa}. 
We express it in terms of $x$, which is related to the cusp angle by $x=e^{i \phi}$.
We have, for example,
\begin{align}\label{cusp1loop}
\Gamma^{(1)}_{\rm cusp}(x) = \frac{1-x}{1+x} (-\log x) \,.
\end{align}
The RHS of eq. (\ref{cusp1loop}) is a CM function of $x \in (0,1)$. This follows from the fact that both factors are CM . 
Similarly, we were able to prove recursively that 
$(-1)^{L+1}  \Gamma_{\rm cusp}^{(L)}$ is a CM function of $x \in (0,1)$. 
%\pr{We initially also checked for the CM property in the angle $\phi$. While we found numerical evidence for the property in $\phi \in (0,\pi)$ up to two loops, at three loops the first derivative changes sign between $1<\phi<3$, so that the property is lost.
%This highlights that the observed CM property in $x$ is non trivial. 
%} 

We note that $\Gamma_{\rm cusp}(x)$ is closely related to the logarithm of the Coulomb branch amplitudes discussed under {\it 1.}, as the latter contain the cusp anomalous dimension in a Regge limit \cite{Henn:2010bk}, where $v\to 0$, in which the variables are matched according to $u=4 x/(1-x)^2$. 
This may suggest that it could be interesting to study CM properties of the cusp anomalous dimension in the variables $u$. We leave this question for future work.

The above examples were 
from planar sYM.
The CM property is however not limited to quantities in this theory. We found it also in the three-loop angle-dependent cusp anomalous dimension in QCD \cite{Grozin:2014hna}, and in the four-loop QED one \cite{Bruser:2020bsh}.
A comment is due regarding their numerical values. In the Euclidean region $x \in (0,1)$, they diverge logarithmically at zero, $\Gamma_{\rm cusp}(x) \sim - \Gamma^{\infty}_{\rm cusp} \log x$, while they vanish by definition at $x=1$. 
Interestingly, as already noted in references \cite{Grozin:2014hna,Bruser:2020bsh} (see also \cite{Kidonakis:2016voy}), if one normalizes the functions so that they have the same small $x$ asymptotics, their graphs are extremely similar, as shown in Fig.~\ref{fig:cusp}. (However, numerically, they differ by several per cent.) 
The remarkable similarity of the plots in that Figure suggests to us that complete monotonicity, together with physical input, could be useful for numerically approximating $\Gamma_{\rm cusp}(x)$.

\section{Summary and Outlook}
 \label{sec:summary}
 
In this work, we have presented evidence that completely monotonic functions play an important role in scattering amplitudes, and in other quantities in QFT.

We explained how complete monotonicity is motivated by Positive Geometry, and in the case of observables given governed by polytopes, follows from Choquet's theorem. In more general cases, the question of complete monotonicity is closely related to the existence of a dual Amplituhedron, which at present is conjectural.

We presented heuristic evidence, based on formulas for multi-loop scattering amplitudes and related quantities available in the literature, for non-trivial CM properties for a number of observables. 
As we discussed, these checks depend on the choice of variables. There are a number of natural variable choices known in the literature for scattering amplitudes, such as e.g. the momenta defining them, Mandelstam invariants, momentum twistors, etc. 
While the CM properties we found in this work are non-trivial, it could be that they are accidental, in the sense that they are not true at some higher loop order, and that the true CM properties, if they are present, hold in a different set of variables.

We also reported on a number of cases where we were able to give elementary proofs of complete monotonicity via  integral representations, from which the infinite number of positivity properties are manifest. This allowed us to show the CM property for scalar Feynman integrals in their Euclidean region, as well as for certain Euler type integrals.

Our findings open up several research directions:

 {\it 1. Developing mathematical proofs for special functions from QFT.}
A pressing issue is to systematically develop methods for proving or disproving the CM property, for the relevant cases of special functions that appear in QFT. This would allow us to go from numerical evidence to rigorous statements about the CM property.

 {\it 2. Exploring mathematical data to find out in which cases complete monotonicity holds.} 
 We are in the fortunate situation to have access to a wealth of analytically known scattering amplitudes, such as for example at higher loops \cite{Dixon:2023kop}, for amplitudes with helicity configurations beyond MHV \cite{Arkani-Hamed:2014dca}, for higher-point Wilson loops with Lagrangian insertions \cite{Chicherin:2022bov}, as well as for various QCD scattering amplitudes.
 It is exciting to study which of these functions have hidden CM properties.

  {\it 3. Relating complete monotonicity to Positive Geometry.} 
  As we argued in subsection \ref{sec:choquet}, Choquet's theorem, cf. eq. (\ref{choquet1}), could lead to a proof of the complete monotonicity for suitable quantities related to positive geometries.
  Along the way, it informs us about a choice of natural variables and region in which to find the CM property. 
An interesting case in point are functions associated to a positive geometry built from linear inequalities, namely the Aomoto polylogarithms \cite{Aomoto_1982}, which, see \cite{Arkani-Hamed:2017tmz} and Appendix B of \cite{Arkani-Hamed:2017ahv}.
In this case, there is a straightforward notion of dual geometry, and hence the conditions of the theorem are satisfied. 
 In order to apply the same logic to the Amplituhedron, it is necessary to extend this to the case of non-linear defining equations.

  {\it 4. Exploring connections to analyticity and unitarity.} 
  {In section \ref{sec:subsectionanalytcSmatrix}, we explained how, based on certain assumptions of analyticity and unitarity, complete monotonicity follows within a certain kinematic region of four-particle scattering amplitudes. This gives rise to a number of questions for future work: can the region of monotonicity be extended to the full Euclidean region? Is there a more direct connection between spectral representations and Choquet's theorem? What constraints can one obtain for multi-particle scattering amplitudes?}

 {\it 5. Harnessing implications of complete monotonicity.} The combination of positivity and convexity has proven to be a successful recipe in physics. This is well appreciated in the context of the conformal field theory and S-matrix bootstrap programs \cite{Simmons-Duffin:2016gjk,Kruczenski:2022lot}, where 
  these principles are used
 to constrain the space of allowed theories. We expect that the knowledge of an infinite number of positivity constraints and convexity of the space of CM functions may be useful for numerical approximations or bootstrap approaches. Just to give one example,
 the additional information we provide may be used to improve the method proposed in \cite{Zeng:2023jek} for numerically bootstrapping Feynman integrals.
  Another direction is to combine positivity with recent machine learning approaches to the symbol bootstrap \cite{Cai:2024znx}. Finally, it would be interesting to explore implications for analytically continued kinematic regions, which are relevant to phenomenological applications of scattering amplitudes. We expect the concept of positive real functions \cite{PRF} to be useful in this regard.

\section*{Acknowledgments}
It is a pleasure to thank Nima Arkani-Hamed, Lance Dixon, Yifei He, Martín Lagares, Elia Mazzucchelli, Sebastian Mizera, Alessandro Podo, Giulio Salvatori, Bernd Sturmfels, and Jaroslav Trnka for discussions.
We also thank Zhenjie Li, Jungwon Lim, Chenyu Wang, and Qingling Yang for help with numerical computations, and Yang Zhang for correspondence.
Funded by the European Union (ERC, UNIVERSE PLUS, 101118787). Views and opinions expressed are however those of the authors only and do not necessarily reflect those of the European Union or the European Research Council Executive Agency. Neither the European Union nor the granting authority can be held responsible for them.

\bibliographystyle{utphys} 
\bibliography{ff.bib}

\end{document}